\documentclass[twocolumn,prb]{revtex4}
\usepackage{amsmath}
\usepackage{amsfonts}
\usepackage{amssymb}
\usepackage{graphics}
\usepackage{float}
\usepackage{graphicx}
\usepackage{epstopdf}
\usepackage[compact]{titlesec}
\usepackage{natbib}
\usepackage{threeparttable}
\usepackage[titletoc,title]{appendix}
\usepackage{lipsum}
\begin{document}

\title{Gate-Voltage Tunability of Plasmons in Single and Multi-layer Graphene Structures: Analytical Description and Concepts for Terahertz Devices}
\author{Shaloo Rakheja$^{1}$}
\email{shaloo.rakheja@nyu.edu}
\email{parijats@bu.edu}
\author{Parijat Sengupta$^{2}$}
\affiliation{$^{1}$ Electrical and Computer Engineering, New York University, New York, NY, 11201. \\
$^{2}$ Photonics Center, Boston University, Boston, MA, 02215.}

\begin{abstract}
The strong light-matter interaction in graphene over a broad frequency range has opened up a plethora of photonics applications of graphene. The goal of this paper is to present the voltage tunability of plasmons in gated single- and multi-layer graphene structures. Device concepts for plasmonic interconnects and antennas and their performance for THz communication are presented. For the first time, the role of gate voltage and the thickness of the gate dielectric on the characteristics of plasmon propagation in graphene are quantified by accounting for both the interface trap capacitance and the quantum capacitance. The gate voltage serves as a powerful knob to tweak the carrier concentration and allows building electrically reconfigurable terahertz devices. By optimizing the gate voltage to maximize the plasmon propagation length in a gated multi-layer graphene geometry, we derive simple scaling trends that give intuitive insight into device modeling and design. 
\end{abstract}
\maketitle

\section{Introduction}
The two-dimensional material graphene, which is a layer of carbon atoms arranged in a honey-comb lattice, exhibits strong light-matter interaction over a broad frequency spectrum from the far infrared to the ultraviolet~\cite{bao12,yan12,avouris14}. The tunability of the density-of-states in graphene along with its excellent transport properties reflected in a high carrier mobility provide a path for graphene photonic applications such as quantum optics~\cite{wang09}, photo-voltaics~\cite{liu08}, photo-detectors~\cite{mueller10,xia09}, and biological sensing~\cite{he10,lu09}. The unique optical properties of graphene originate from a combination of its two-dimensional nature and gapless electronic spectrum.
In the optical frequency range, the universal dynamical conductivity in graphene\footnote{Note that the universal dynamical conductivity is different from the universal DC conductivity, which is given as $\sigma_{min}^{DC} = 4e^2/(\pi h)$.} is $\sigma_{min} = \pi e^2/2h$, where $e$ is the elementary charge, and $h$ is the Planck's constant.  This provides mono-layer graphene with a transparency $T = (1+2\pi\sigma_{min}/c)^{-2} \approx 1-\pi\alpha  = 0.977$, where $\alpha = e^2/(\hbar c)$ is the fine-structure constant, $\hbar$ is the reduced Planck's constant, and $c$ is the speed of light. Hence, graphene has been sought after as a transparent electrode in solar cell and other flexible electronics applications \cite{jena13}. 

Further, the excitation of plasmons, which are the quantized collective oscillations of charged carriers at the interface between graphene and dielectric, are responsible for several interesting optical attributes of graphene \cite{kumada13}. Plasmons lead to the dramatic alteration in optical absorption features in graphene and can be localized within small regions compared to the wavelength of the incident radiation. In graphene, the propagation length of plasmons can be several micrometers and their propagation velocity has a lower bound of $v_f$/2, where $v_f = 8\times 10^5$ m/s is the Fermi velocity of the Dirac fermions in graphene. The long plasmon lifetime and their very high propagation velocity make graphene an ideal platform for implementing plasmonic waveguides for on-chip communication and ultra-broadband antennas for wireless communication. To design photonic devices using graphene, analytical descriptions of optical properties in graphene prove useful.

We begin with a discussion of the plasmon dispersion relationship for both ungated and gated graphene structures in Section II. This section details the key differences in the propagation of surface plasmons in graphene upon the application of a gate voltage. In Section III, we present the functional relationship between the Fermi level (carrier concentration) in graphene and the gate voltage using a circuit model to account for trap charges due to dangling bonds at the interface between graphene and the substrate. 
In Section IV, we combine the results from Sections II and III to quantitatively predict figures of merit for graphene plasmonic waveguides and antennas for wireless THz communication. The paper concludes in Section IV with a summary of the results presented in the paper and an outlook on the future of graphene plasmonics. We also supplement the work with three appendices that detail out the specific mathematical calculations underlying the theoretical foundations of the paper.

\vspace{0.35cm}
\section{Plasmon dispersion relation in single- and multi-layer graphene}
\vspace{0.35cm}
The plasmon dispersion relationship in graphene is derived from finding the roots of the real part of the dielectric constant, $\epsilon_{E_f}$, given as~\cite{hwang_plasmon07,luo13}
\begin{eqnarray}
\epsilon_{E_f}(q,\omega) = \epsilon-v_q\Re{\Pi_{E_f}(q,\omega}) = 0,
\label{eq_dielectric_function}
\end{eqnarray}
where $E_f$ is the Fermi level in graphene, $\epsilon$ is the permittivity of the surrounding media, $v_q = e^2/2\epsilon_0q$ is the Coulomb interaction term, and $\Pi_{E_f}(q,\omega)$ is the polarization function of doped graphene. In the above equation, $q$ denotes the wave-vector, and $\omega$ is the frequency. 
Since $v_q > 0$, (\ref{eq_dielectric_function}) has a solution only when $\Re{\Pi_{E_f}(q,\omega}) > 0$. It can be shown that the inequality is satisfied only when $\omega > v_f q$. In violation of this condition, one arrives at electron-hole continuum in the regime of single-particle excitation, which precludes the existence of plasmons \cite{sasaki12}.
Ignoring the non-local effects and in the long wavelength limit ($q \rightarrow 0$) allows approximating the polarization function in doped graphene according to 
\begin{eqnarray}
\Re{\Pi_{E_f}(q,\omega})= \frac{|E_f|}{\pi}\left(\frac{q}{\hbar\omega}\right)^2.
\end{eqnarray}

In this case, a simplified plasmon dispersion relation in graphene is given as~\cite{tonylow}
\begin{eqnarray}
\omega(q) = \frac{1}{\hbar}\sqrt{\frac{e^2|E_f|q}{2\pi\epsilon_0\epsilon}}.
\label{eq_omega_q_simple}
\end{eqnarray}

Alternatively, the plasmon dispersion relationship can be derived by using Maxwell's equations and applying appropriate boundary conditions for a TM-polarized electromagnetic wave interacting with graphene embedded in a dielectric environment within the non-retarded regime where $q >> \omega/c$. More details on the derivation of the plasmon dispersion relationship for TM-polarized plasmon modes in graphene are given in Appendix A. The plasmon dispersion relation in graphene is distinctly different from other 2D materials where the dependence of $\sqrt{q}$ is certainly obtained, but the dependence on $\sqrt{E_f}$ is unique to graphene because of the presence of massless Dirac fermions at Brillouin zone edges $ K $ and $ K^{'} $. Using (\ref{eq_omega_q_simple}), the plasmon propagation velocity, $v_g = \partial \omega/\partial q $ is given as
\begin{eqnarray}
v_g(q) = \frac{1}{\hbar}\sqrt{\frac{e^2|E_f|}{8\pi\epsilon_0\epsilon q}}.
\label{eq_vg_simple} 
\end{eqnarray}
Noting that the assumption for the derivation of the dispersion relation in (\ref{eq_omega_q_simple}) is that $\omega > v_f q$, it can be shown using (\ref{eq_omega_q_simple}) and (\ref{eq_vg_simple}) that the lower bound on the propagation velocity of graphene plasmons is $v_f/2$. 

The propagation length of plasmons in graphene is given as 
\begin{eqnarray}
L_{prop} = \frac{v_g}{\Gamma},
\label{eq_propagation_length}
\end{eqnarray}
where the factor $\Gamma$ is a phenomenological parameter that characterizes the electron scattering rate in graphene~\cite{rakheja2015tuning}. An estimate of $\Gamma$ can be obtained from the D.C. relaxation time, which arises mainly from scattering due to intrinsic phonons and the charged impurities of the substrate. In this work, we consider both scattering events, which are carrier concentration dependent, to obtain the net electron scattering rate in graphene. Details of various electron scattering times in graphene are noted in Appendix B, while the modulation of the carrier concentration with an external gate voltage is discussed in Section III.

\vspace{0.35cm}
\subsection{Impact of metal top gate on plasmon dispersion in graphene}
\begin{figure}[h!]
\centering
\includegraphics[scale = 0.4]{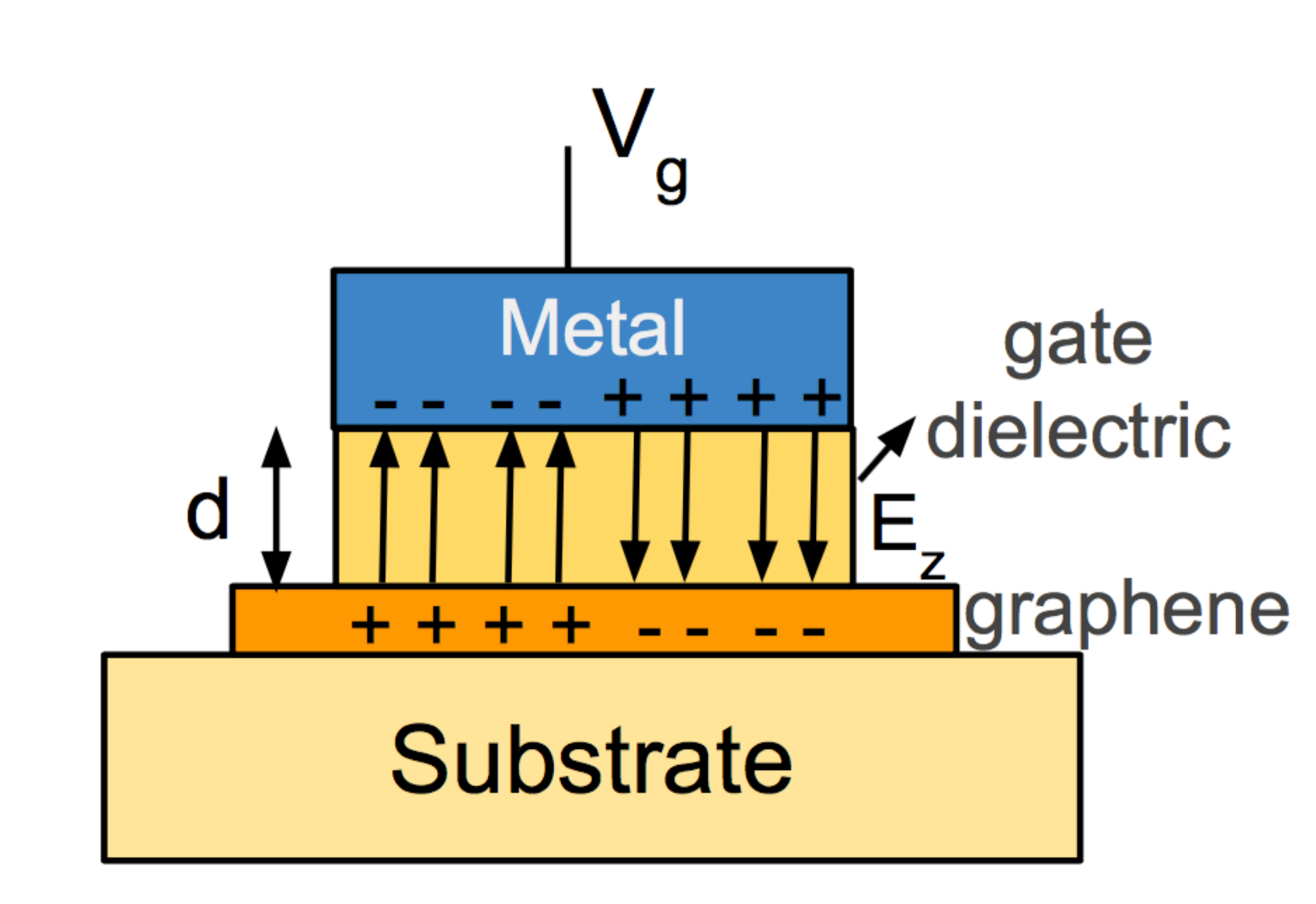} 
\caption{Gated graphene structure with a top gate with a bias $V_g$ leads to altered plasmon dispersion relation.}
\label{fig_gated_graphene}
\end{figure}

\vspace{0.35cm}
When a metal top gate is placed on the graphene sheet with a gate dielectric as shown in Fig. \ref{fig_gated_graphene}, the electric field distribution is modified such that the electric field has only the $z-$ component in the direction perpendicular to the graphene sheet as noted in the figure. This is in stark contrast to the electric field distribution in ungated graphene structure, where electric field has components in both $z - $ and $x - $ directions, where the $x - $ direction is along the length of the graphene ribbon and coincides with the direction of plasmon propagation. Upon application of the gate voltage, the plasmon dispersion relation in graphene is modified according to 
\begin{eqnarray}
\omega(q) = \sqrt{\frac{\sigma_0\Gamma}{\epsilon_0\epsilon}}\sqrt{\frac{q}{1+\coth{qd}}},
\label{eq_omega_metalgate}
\end{eqnarray}   
where $\sigma_0$ is the D.C. conductivity of the graphene sheet and is dependent on the Fermi level in graphene, $\Gamma$ is the electron scattering rate, and $d$ is gate dielectric thickness. The plasmon dispersion relation given above was first discussed by Nakayama and co-workers in the context of plasmons in 2D electron gas~\cite{nakayama74}, where the conductivity in (\ref{eq_omega_metalgate}) stands for the conductivity of the 2D electron gas. Later,  
the same dispersion relationship for plasmons in gated graphene structures was derived for graphene in~\cite{ryzhii06}, \cite{ryzhii07} and discussed in detail in~\cite{sasaki14}.

Using the random-phase approximation in the long wavelength limit ($q$ $\rightarrow$ 0), the frequency-dependent intra-band conductivity of mono-layer graphene can be expressed as \cite{falkovsky08}
\begin{eqnarray}
\sigma_{intra}(\omega) = \frac{e^2 \omega}{i\pi\hbar}\int_0^{\infty}{\frac{dE |E|}{\omega^2}\frac{df_0(E)}{dE}},
\end{eqnarray}
where $f_0(E) = \left(1+exp\left((E-E_f)/k_BT\right)\right)^{-1}$ is the Fermi function.
Carrying out the integration, $\sigma_{intra}(\omega)$, can be written more simply as 
\begin{eqnarray}
\sigma_{intra}(\omega) = i\frac{2e^2}{\pi\hbar^2}\frac{k_BT}{\left(\omega+i\Gamma\right)}\ln\left[2\cosh{\frac{|E_f|}{k_BT}}\right],
\end{eqnarray}
where we have replaced $\omega$ with $\omega+i\Gamma$ to account for the electron scattering rate in graphene. Further, using the simplification for the function $\cosh{(x)}$, we can show that the intra band conductivity reduces to the form:
\begin{eqnarray}
\sigma_{intra}(\omega) = i\frac{2e^2}{\pi\hbar^2}\frac{k_BT}{\omega+i\Gamma}\left[\frac{E_f}{2k_BT}+\ln{\left(1+e^{-E_f/k_BT}\right)}\right].
\end{eqnarray}
The D.C. conductivity from the above equation is found by setting $\omega << \Gamma$ and is given as 
\begin{eqnarray}
\sigma_0 = \frac{2e^2}{\pi\hbar^2}\frac{k_BT}{\Gamma}\left[\frac{E_f}{2k_BT}+\ln{\left(1+e^{-E_f/k_BT}\right)}\right].
\end{eqnarray}
In the zero temperature limit, T $\approx$ 0, which is a good approximation for heavily-doped graphene ($E_f >> k_BT$), 
\begin{eqnarray}
\sigma_0 = \frac{e^2|E_f|}{\pi\hbar^2\Gamma}.
\label{eq_sigma_simple}
\end{eqnarray}
Using this definition of $\sigma_0$ in (\ref{eq_omega_metalgate}), $\omega(q)$ can be simplified as
\begin{eqnarray}
\omega(q) = \frac{1}{\hbar}\sqrt{\frac{e^2|E_f|}{\pi\epsilon_0\epsilon}}\sqrt{\frac{q}{1+\coth{qd}}}.
\label{eq_omega_metalgate2}
\end{eqnarray} 
In the limit $qd \rightarrow \infty$, (\ref{eq_omega_metalgate2}) gives the same results as (\ref{eq_omega_q_simple}). In the limit $qd << 1 $, $\omega(q)$ in (\ref{eq_omega_metalgate2}) approaches
\begin{eqnarray}
\omega(q) = \frac{1}{\hbar}\sqrt{\frac{e^2|E_f|d}{\pi\epsilon_0\epsilon}}q, (qd << 1).
\end{eqnarray}  
As evident from the above equation, in the gated graphene structure, plasmon-dispersion relation depends linearly on the wave-vector, while in the ungated graphene structure the plasmon dispersion relation varies as $\sqrt{q}$. That is, in gated graphene structures, plasma waves exhibit a sound-wave-like dispersion relation. The frequency of these plasma waves can be tuned by gate voltage within a fairly wide frequency band.

\begin{figure}[h!]
\centering
\includegraphics[width=2.8in]{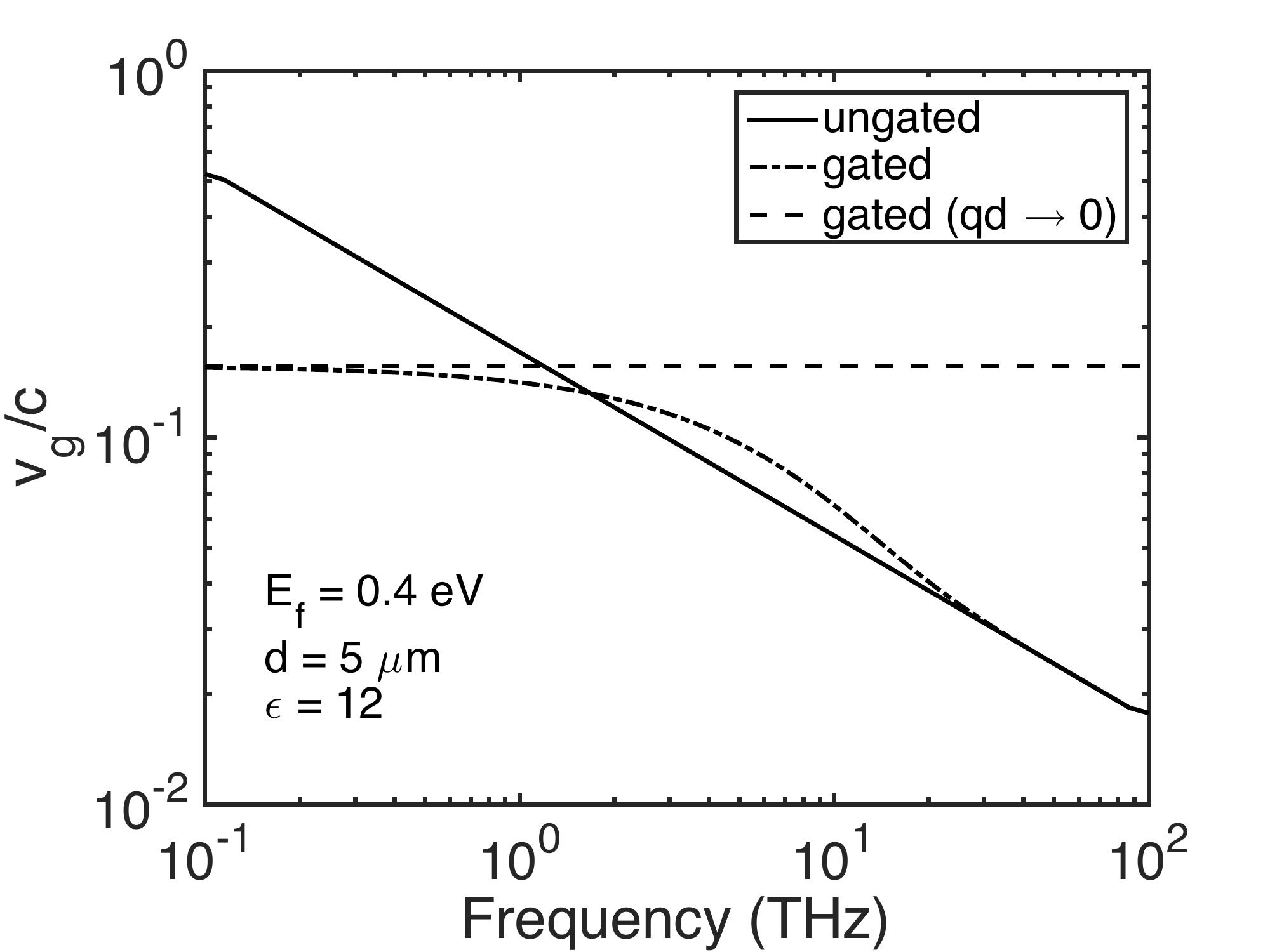} 
\caption{Plasmon propagation velocity in mono-layer graphene normalized to the speed of light versus frequency of operation. Three cases are considered: (i) ungated graphene structure for which $v_g$ is described by (\ref{eq_vg_simple}), (ii) gated graphene structure in the general case for which $v_g$ described by (\ref{eq_vg_gated_general}), and (iii) gated graphene structure for which $qd << 1$ such that $v_g$ is independent of the wave vector and, therefore, the frequency of operation as described by (\ref{eq_vg_gated_lowlimit}). Here, $E_f$ denotes the Fermi level in graphene, $d$ is the thickness of the gate dielectric, and $\epsilon$ is the average dielectric of the surrounding media.}
\label{fig_vg_vs_freq_difflimits}
\end{figure}

The difference between the dispersion relations in gated and ungated graphene structures translates to a significant difference in their plasmon propagation velocity. Using (\ref{eq_omega_metalgate}), the plasmon propagation velocity, $v_g$, is given as
\begin{eqnarray}
v_g = \frac{d}{2}\sqrt{\frac{\sigma_0\Gamma q}{\epsilon\epsilon_0\left(1+\coth(qd)\right)}} \left[\frac{1}{qd}+\frac{\text{csch}{qd}^2}{1+\coth{qd}}\right].
\label{eq_vg_gated_general}
\end{eqnarray}   
In the limit $qd << 1$, the plasmon propagation velocity approaches the limit
\begin{eqnarray}
v_g = \sqrt{\frac{\sigma_0\Gamma d}{\epsilon_0\epsilon}}.
\label{eq_vg_gated_lowlimit}
\end{eqnarray}
Note that, in this case, the plasmon propagation velocity is independent of the wave-vector and is a constant that depends on the material properties and the geometric dimension $d$. This behavior is significantly different from that described in (\ref{eq_vg_simple}), where the propagation velocity scales as $\sqrt{q}$, a characteristic feature of 2D electron gas. As shown in Fig. \ref{fig_vg_vs_freq_difflimits}, the plasmon propagation velocity for a gated graphene structure converges to that for the ungated graphene structure in the limit $qd \rightarrow \infty$. In the other extreme when $qd << 1$, the propagation velocity is simply given as (\ref{eq_vg_gated_lowlimit}), and in this case, the propagation velocity is a constant independent of the wave-vector. 

\begin{figure}[h!]
\centering
\includegraphics[width=2.8in]{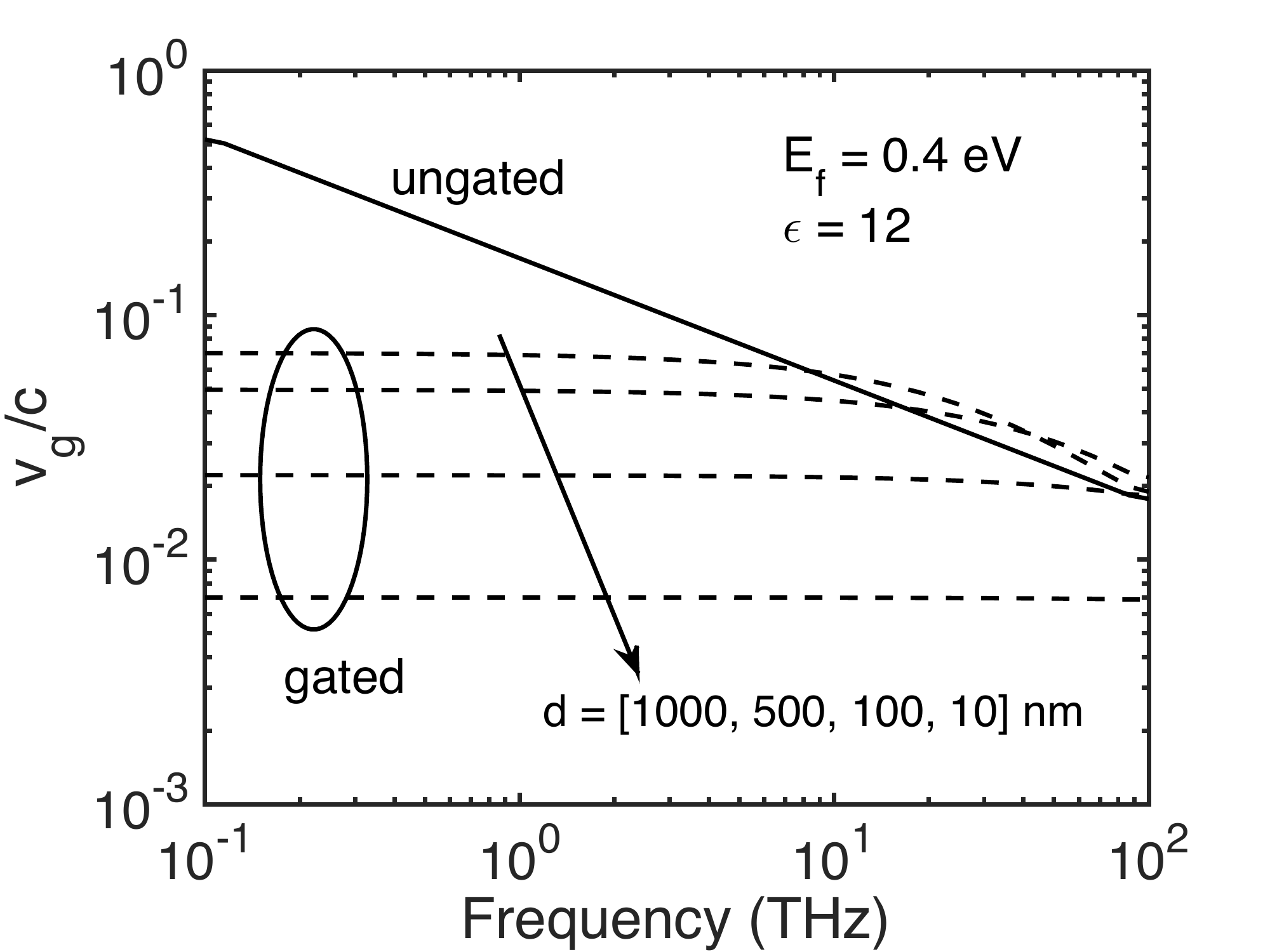} 
\caption{Plasmon propagation velocity versus frequency of operation for different thicknesses of the gate dielectric, $d$.}
\label{fig_vg_diff_distance}
\end{figure}

It must also be noted that in the limit $qd << 1$, the propagation velocity of plasmons is lower than the propagation velocity of ungated plasmons. For a fixed Fermi-level, in this regime, the plasmon propagation velocity scales as $\sqrt{d}$, such that the plasmon propagation velocity in gated graphene structures with very thin gate dielectrics may be significantly lower than the plasmon propagation velocity in ungated graphene structres as shown in Fig. \ref{fig_vg_diff_distance}. Despite the lower propagation velocity, the advantage of the gated graphene structure is to provide dynamic tunability of the plasmon characteristics in graphene depending on the required circuit functionality. This ``reconfigurability'' option is a fundamental advantage of graphene plasmonics over metal-based plasmonics. As discussed in Section III, the gate voltage is used to tune the carrier concentration and, therefore, the Fermi level in the graphene sheet.

We would like to note that the plasmon dispersion relation derived for mono-layer graphene in the THz frequency will  be preserved in the case of multi-layer graphene as long as the plasma wavelength is much larger than the thickness of the sheet \cite{nakayama74}. However, the carrier concentration  must include contribution from each layer in the multi-layer stack. For perfectly coupled layers considered in this work, $\sigma_{multi} = N_{layer}\sigma_{mono}$, where $\sigma_{multi}$ ($\sigma_{mono}$) is the multi- (mono-) layer conductivity of graphene, and $N_{layer}$ is the number of layers in the multi-layer stack.

\vspace{0.35cm}
\section{Modulation of Fermi level via gate voltage in graphene}
\begin{figure*}
\centering
\includegraphics[width=5in]{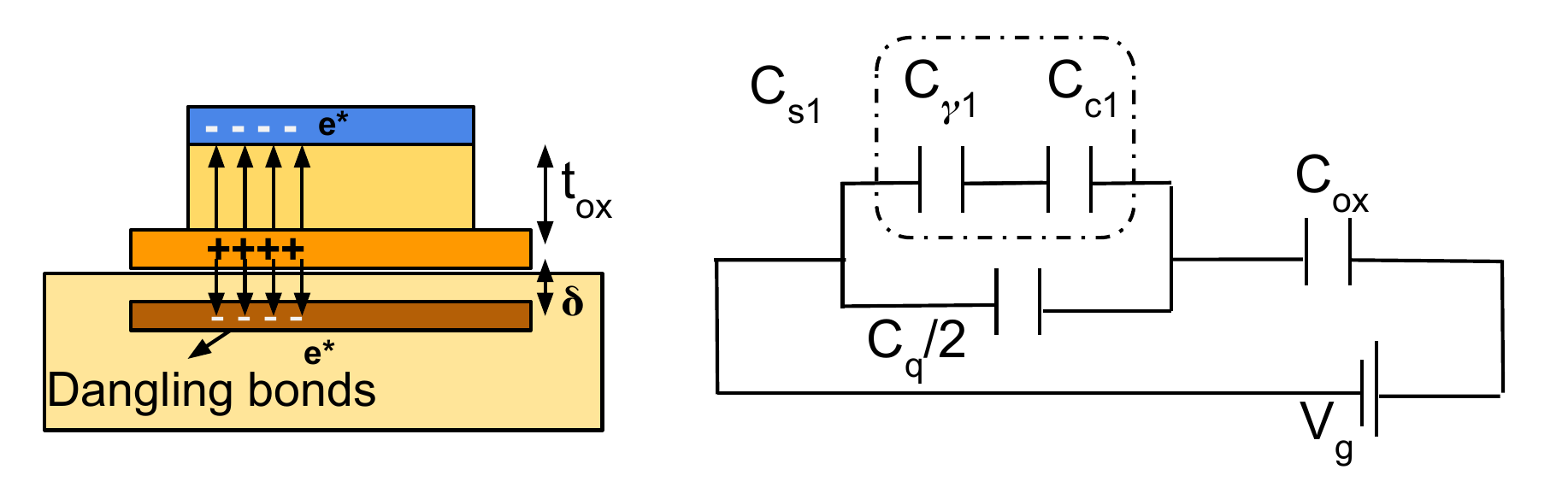} 
\caption{Schematic of a graphene sheet on a substrate with dangling bonds. Equivalent circuit model that acounts for interface traps due to dangling bonds is also shown. $C_{c1}$ and $C_{\gamma_1}$ are the electrostatic and quantum capacitance, respectively, of the interface traps. $C_q$ is the quantum capacitance of the graphene sheet.}
\label{fig_monographene_eq_ckt}
\end{figure*}

\vspace{0.35cm}
To determine the modulation of carrier concentration, $n_s$, in the graphene sheet due to the gate voltage, $V_g$, the capacitance voltage network shown in Fig. \ref{fig_monographene_eq_ckt} is used. As shown, the equivalent circuit accounts for the capacitance due to interface traps created due to dangling bonds at the interface between graphene and the substrate. From the equivalent circuit diagram, the relationship between the Fermi level, $E_f$, in the graphene sheet and the gate voltage, $V_g$, can be expressed as
\begin{eqnarray}
\frac{E_f}{e} = \frac{C_{ox}}{C_{ox}+C_{s1}+1/2C_q}\left(V_{g}-V_{min}\right).
\label{eq_Ef_Vg_relation}
\end{eqnarray}
Here, $C_{s1}$ is given as the series combination of $C_{c1}$ and $C_{\gamma_1}$, where $C_{c1}$ is the electrostatic capacitance of the interface traps with the graphene sheet, and $C_{\gamma_1}$ is the quantum capacitance of the interface traps. $C_{\gamma_1}$ is given as $e^2\gamma_1$ with $\gamma_1$ being the energy-independent density-of-states of the interface traps. Typical values of $\gamma_1 $ are $\approx 5\times 10^{12}$ $eV^{-1}cm^{-2}$ as measured experimentally in \cite{janssen11} and \cite{kopylov10} and also considered by authors in \cite{takase12} to fit their theoretical model to explain the dependence of $n_s$ on $V_g$. 
In the above equation, $V_{min}$ is the Dirac point voltage of the graphene sheet, and $C_q = e^2 dn_s/dE_f$ is the quantum capacitance of graphene.

\begin{figure*}
\centering
\includegraphics[width=5in]{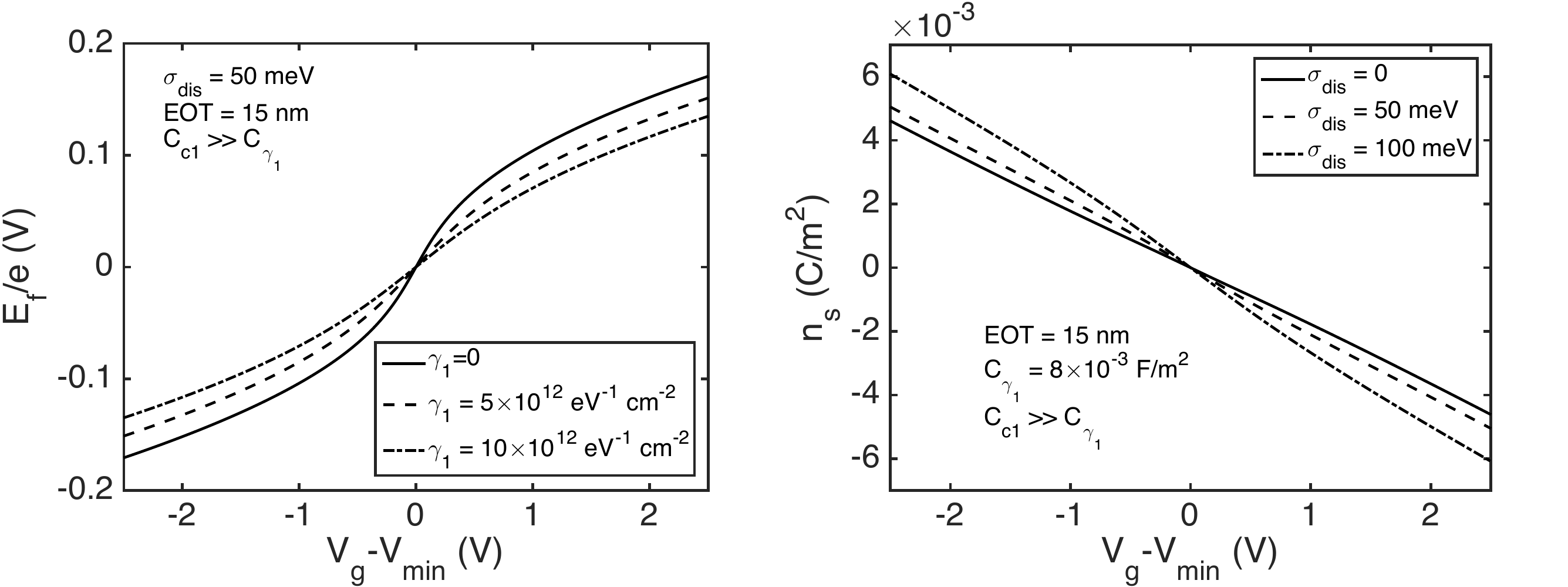} 
\caption{(left) Dependence of Fermi level and (right) sheet carrier concentration on gate voltage. In the left hand side plot, different traces correspond to the different values of the density-of-states, $\gamma_1$ of the interface traps. In the right hand side figure, different traces correspond to the different values of the broadening in the density-of-states, $\sigma_{dis}$, around the Dirac point in graphene.}
\label{fig_Ef_ns_Vg}
\end{figure*}

To relate the carrier concentration, $n_s$, with the Fermi level, $E_f$, in the graphene sheet, the following Fermi-Dirac integrals are used:
\begin{subequations}
\begin{equation}
n_{s}  = |n_{elec}(E_f)-n_{hole}(E_f)|,
\end{equation}
\begin{equation}
n_{elec}(E_f) = \int_0^{\infty}DOS(E)\frac{dE}{1+exp\left(\frac{E-E_f}{k_BT}\right)},
\end{equation}
\begin{equation}
n_{hole}(E_f) = \int_0^{\infty}DOS(E)\frac{dE}{1+exp\left(\frac{E+E_f}{k_BT}\right)},
\end{equation}
\end{subequations}   
where $DOS(E)$ is the 2D density-of-states in graphene. In the presence of electron-hole puddles, the density-of-states is more appropriately given as \cite{rakheja_avs14}
\begin{eqnarray}
DOS(E) - \frac{2}{\pi(\hbar v_f)^2}[\frac{2\sigma_{dis}}{\sqrt{2\pi}}exp\left(-\frac{E^2}{2\sigma_{dis}^2}\right) \nonumber \\
+E\times erf\left(\frac{E}{\sigma_{dis}\sqrt{2}}\right)],
\label{eq_DOS_disorder}
\end{eqnarray}
where $\sigma_{dis}$ is the broadening in eV of the DOS around the Dirac point. Using (\ref{eq_Ef_Vg_relation})-(\ref{eq_DOS_disorder}), we show the dependence of $E_f$ and $n_s$ on the gate voltage in Fig. \ref{fig_Ef_ns_Vg}. As expected, with an increase in the interface traps DOS, the dependence of $E_f$ on the gate voltage is weakened, since for the same gate voltage a large amount of capacitance comprising of the parallel combination of $C_{s1}$ and $1/2C_q$ will need to be charged reducing the Fermi level in the graphene sheet. An increase in the broadening in the DOS around the Dirac point enhances the net sheet charge concentration in the graphene sheet.   

\vspace{0.35cm}
\section{Gate-voltage-controlled plasmon propagation characteristics in mono- and multi-layer graphene ribbons and implications for THz wireless communication}
\vspace{0.35cm}
In this section, we quantify the limits of plasmon-based on-chip communication by utilizing the gate-voltage tunability of the plasmon dispersion relationship in both mono- and multi-layer graphene structures. We also present a preliminary analysis of the frequency response of patch antennas with graphene targeted toward several interesting communication and radar applications in the THz band. Graphene patch antennas have previously been discussed in \cite{josep13}. However, prior work focuses on narrow graphene ribbons where the Fermi level is only chemically tunable via doping the graphene sheet. Also, earlier works such as \cite{rakheja_drc14} on graphene plasmonic on-chip interconnect analysis focus only on mono-layer and chemically-doped graphene ribbons.
Two useful figures of merit are discussed in detail in the next sub-sections (i) plasmon propagation velocity that determines the latency of high-speed on-chip interconnects, and (ii) plasmon propagation length relating the energy per bit for communication and the radiation efficiency of plasmonic antennas. A large propagation length signifies lower ohmic losses in graphene and, therefore, superior radiation efficiency of the antenna. 

\vspace{0.35cm}
\subsection{Propagation characteristics versus gate voltage in 2D graphene}
\begin{figure}[h!]
\centering
\includegraphics[width=3in]{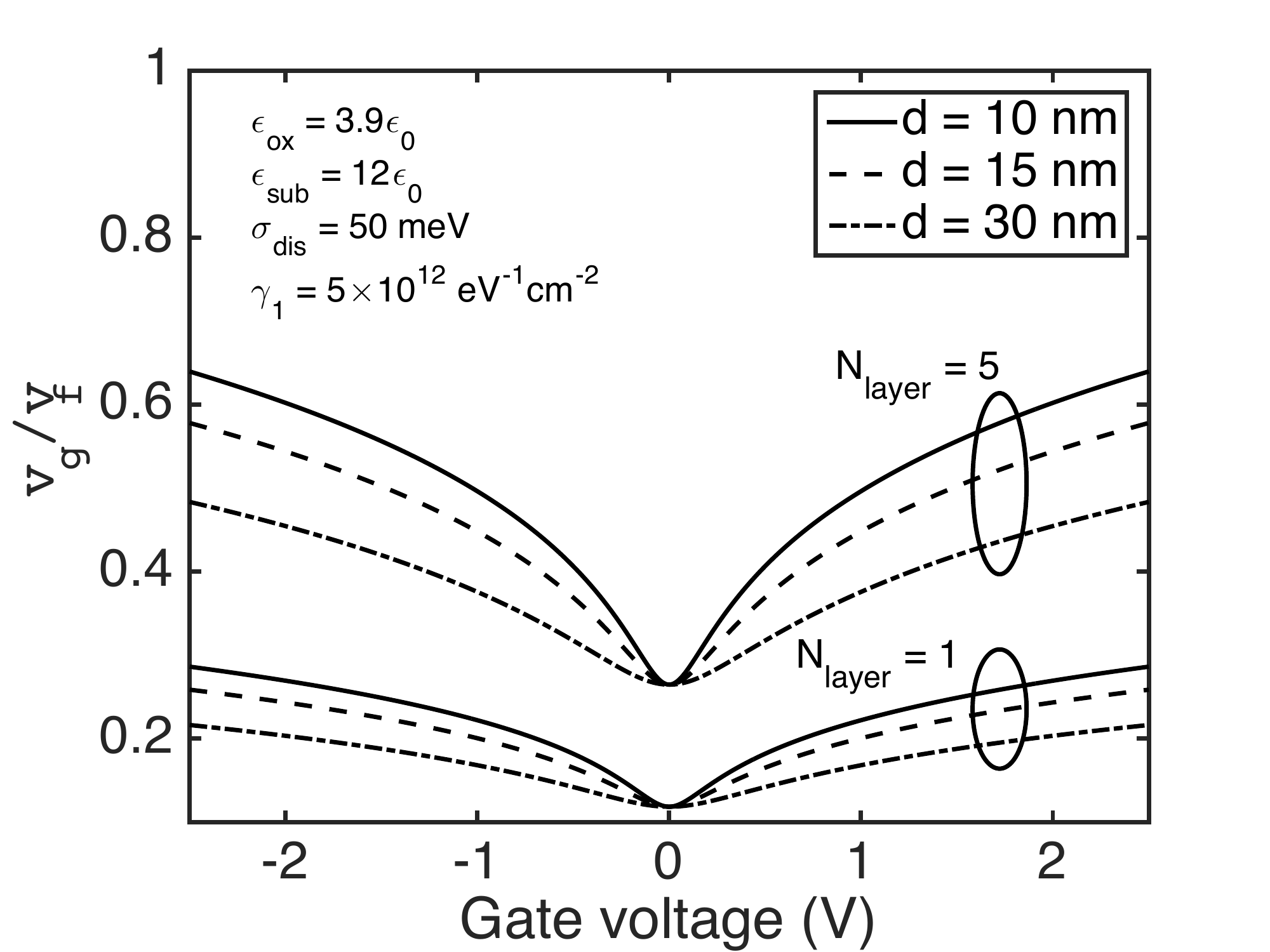}
\caption{Gate voltage impact on the propagation velocity of plasmons in 2D graphene with $N_{layer}$ = 1 (mono-layer) and $N_{layer}$ = 5 (multi-layer) with different values of the gate dielectric thickness.} 
\label{fig_prop_vel_gate_voltage_2D}
\end{figure}

\vspace{0.35cm}
In a gated graphene structure, propagation characteristics of plasmons are tunable by changing the gate voltage. In Fig. \ref{fig_prop_vel_gate_voltage_2D} we show plasmon propagation velocity versus gate voltage for different values of the number of graphene layers and varying dielectric thicknesses. Due to an increase in the 2D conductivity of the graphene sheet with an increase in the number of layers in the multi-layer stack, the propagation velocity is superior for $N_{layer}$ = 5 versus $N_{layer}$ = 1. However, an important difference must be noted between the results in Fig. \ref{fig_prop_vel_gate_voltage_2D} compared to the results in Fig. \ref{fig_vg_diff_distance}. Both figures display an opposite dependence on the parameter $d$, which is the gate dielectric thickness. The reason is that in Fig. \ref{fig_prop_vel_gate_voltage_2D} an increase in $d$ leads to a reduction in the Fermi level in the graphene sheet because of the reduced gate control. This leads to a reduction in the conductivity, $\sigma_0$, which in turn reduces the plasmon propagation velocity. In Fig. \ref{fig_vg_diff_distance}, we had assumed a constant value of $E_f$ in the graphene sheet and $E_f$ and $d$ were adjusted independently of each other. 

\begin{figure}[h!]
\centering
\includegraphics[width=3in]{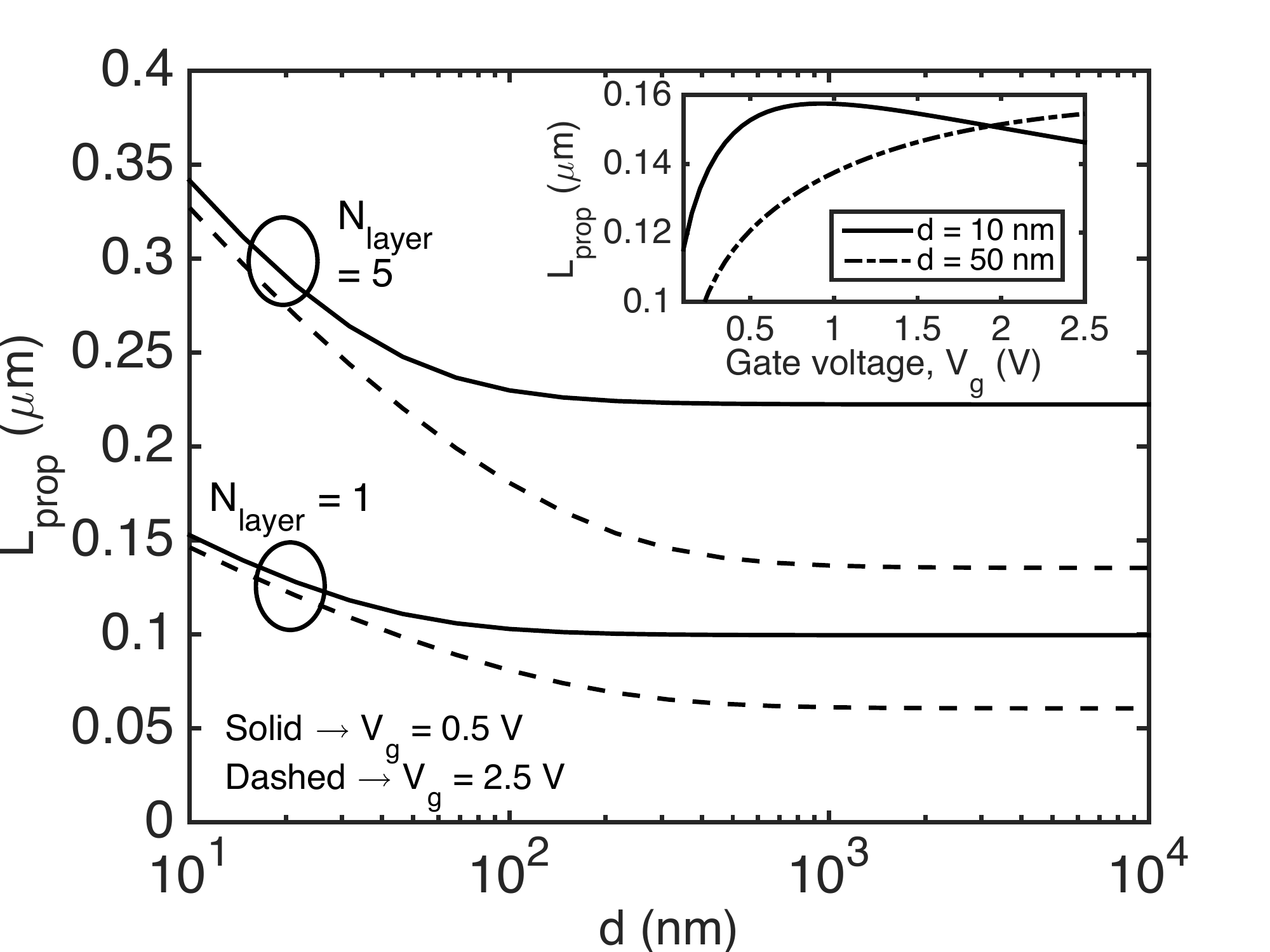} 
\caption{Propagation length versus the gate dielectric thickness. The inset plot shows the dependence of the propagation length on the gate voltage.}
\label{fig_Lprop_vs_product}
\end{figure}

We show the dependence of the plasmon propagation length on the thickness of the top gate for different gate voltages and different number of layers in a gated multi-layer graphene stack in Fig. \ref{fig_Lprop_vs_product}. As the thickness $d$ increases, the propagation length drops and becomes independent of $d$ for very large value of $d$. This is expected since in the limiting case of $qd \rightarrow \infty$, the dependence of $v_g$ on $d$ vanishes. This makes the plasmon propagation length independent of $d$ in the limiting case. For the same value of relaxation rate, $\Gamma$, propagation length improves with the number of layers in the multi-layer stack. It must be noted that while the propagation velocity increases monotonically with an increase in the Fermi level in graphene, the propagation length of plasmons is a non-monotonic function of the Fermi level. This is explained by observing that an increase in $E_f$ enhances both $v_g$ and $\Gamma$ such that the propagation length as defined in (\ref{eq_propagation_length}) exhibits a non-monotonic dependence on $E_f$ and, therefore, on the gate voltage $V_g$ as shown in the inset plot of Fig. \ref{fig_Lprop_vs_product}.

\begin{figure}[h!]
\centering
\includegraphics[width=3in]{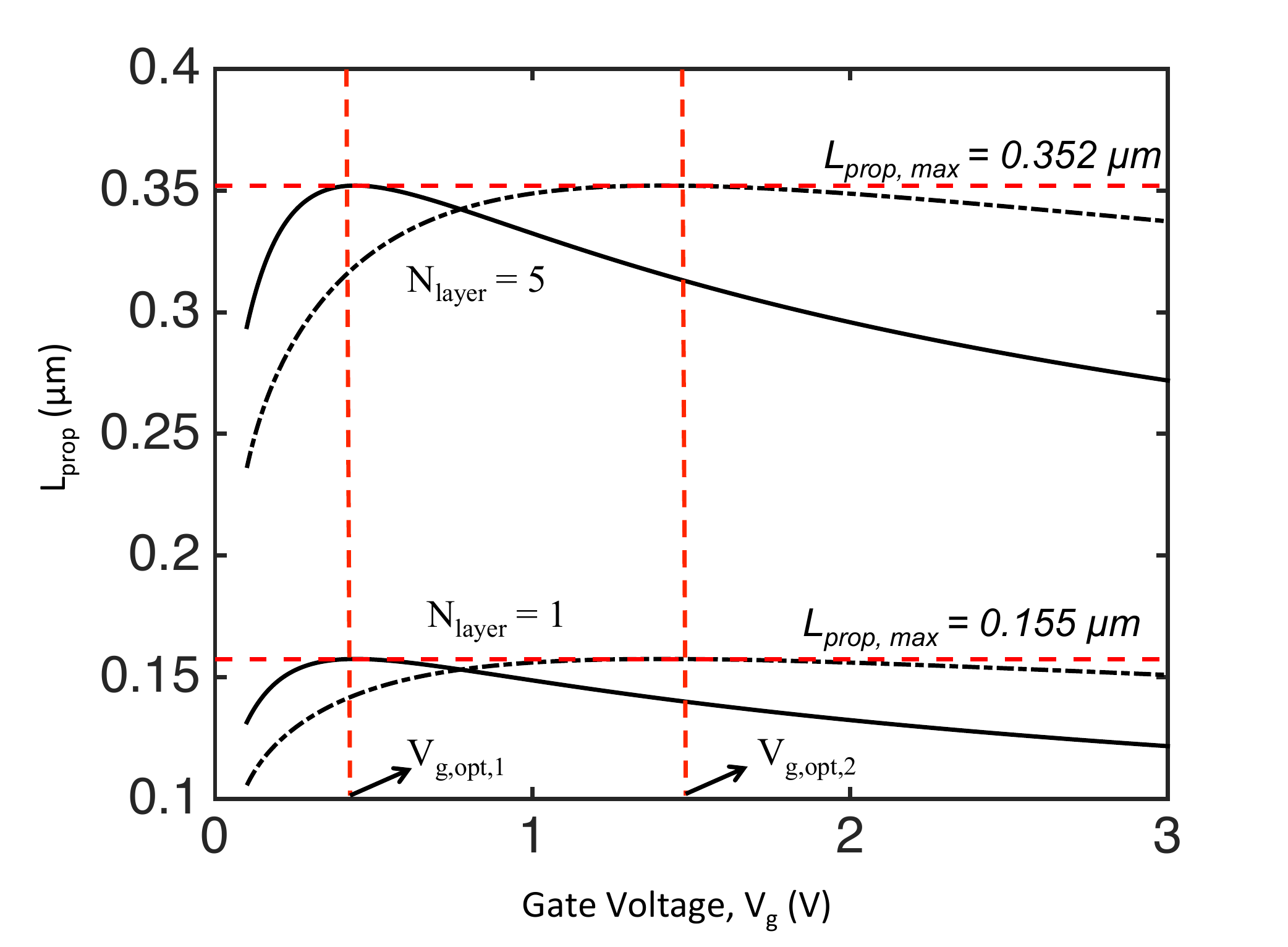} 
\caption{Gate voltage dependence of the propagation length. An optimal gate voltage is identified that maximizes $L_{prop}$ for a given number of layers in the multi-layer stack. The value of the optimal gate voltage is independent of the number of layers. Maximum propagation length is independent of the gate voltage.}
\label{fig_Lprop_vs_Vg}
\end{figure}

As shown in Fig. \ref{fig_Lprop_vs_Vg}, an optimization of the gate voltage to maximize plasmon propagation length reveals two noteworthy features. First, the optimal voltage is independent of the number of layers in the multi-layer stack if the electron relaxation rate is independent of the number of layers. While the optimal gate voltage to maximize the propagation length scales linearly with the value of the gate dielectric thickness, the maximum value of the propagation length stays constant for a given number of layers in the  multi-layer stack. This allows us to identify a simple scaling law for optimizing the gate voltage to obtain maximum plasmon propagation for the graphene structure. The scaling law can be stated as
\begin{eqnarray}
V_{g,opt}(d) = V_{g,opt}(d_0)+\Delta(d-d_0),
\end{eqnarray} 
where $\Delta$ depends on the relative strengths of the different scattering processes in graphene, and $d_0$ is the reference dielectric thickness. Typical value of $\Delta$ $\approx$ (70-100) mV/nm for a impurity concentration of (5-10)$\times 10^{14}$ $cm^{-2}$. The value of $\Delta$ increases with an increase in the impurity concentration. Details of the derivation of the scaling law are provided in Appendix C.  

\vspace{0.35cm}
\subsection{Microstrip antennas with graphene plasmon resonant cavity}
\begin{figure}
\centering
\includegraphics[width=3in]{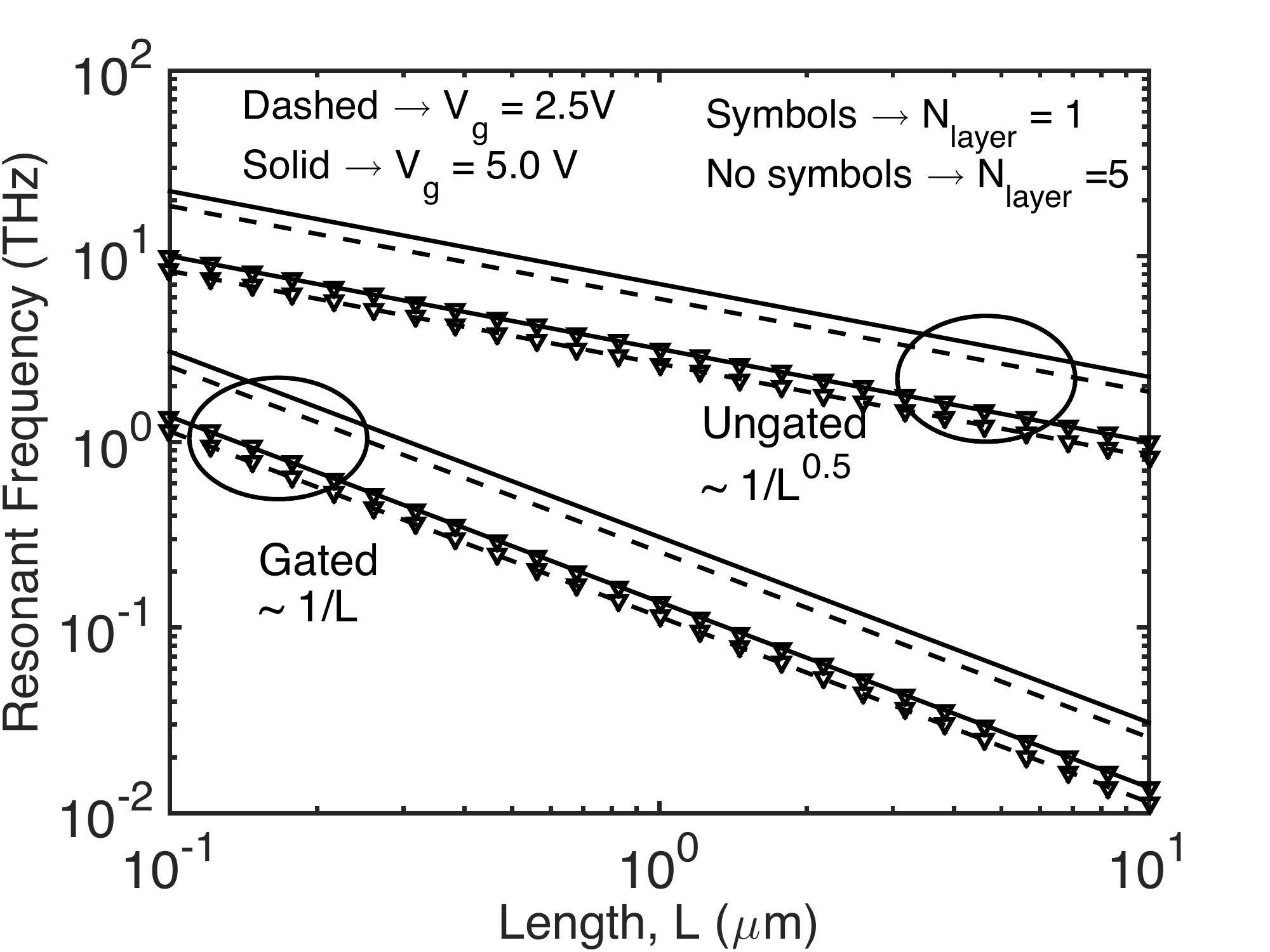}
\caption{Resonant frequency as a function of patch length for gated and ungated graphene microstrip antennas.}
\label{fig_resonant_frequency_2D_vs_length} 
\end{figure}

\vspace{0.35cm}
Microstrip  or patch antennas are used in several communication applications owing to their simplicity, ease of fabrication, and unidirectional radiation. The advantage of graphene to implement high-frequency microstrip antennas is its planar structure and flexibility to be transferred on several different substrates. This eases the integration of graphene within the existing semiconductor nanotechnologies spanning from biological and chemical nanosensor networks to optical interconnects in advanced multi-core architectures. In this work, we model graphene microstrip as a plasmon resonant cavity. As discussed in \cite{jornet13_book}, a resonant cavity of graphene will necessitate the following dimensional constraints to be satisfied
\begin{eqnarray}
L >> W >> h,
\end{eqnarray} 
where $L$ and $W$ are the length and the width of the resonant cavity, while $h$ is the height of the resonant cavity from the substrate. Further, the length of the resonant cavity must be selected for a given resonant frequency response. For a TM plasmon propagation mode, which is the focus of this work, the length $L$ of the graphene resonant patch must satisfy
\begin{eqnarray}
L = m \frac{\lambda_{spp}}{2} = m \frac{\pi}{\Re{q}},
\end{eqnarray}  
where $m = 1, 2, ....$, and $\lambda_{spp}$ is the wave-vector of the propagating plasmon mode, and $q$ is the plasmon wave-vector as discussed in the previous sub-sections. For the first TM propagation mode with $m$ = 1, we plot the resonant frequency versus the length of the graphene sheet for different layers in the graphene multi-layer stack in Fig. \ref{fig_resonant_frequency_2D_vs_length}. Both gated and ungated graphene structures are considered. As seen, the resonant frequency spans from few hundreds of GHz to a few THz depending on the length of the patch and the position of the Fermi level, which has been adjusted using the gate voltage. 
If one were to utilize chemical doping to achieve the desired Fermi level in the graphene sheet such that the plasmon dispersion relation is expressed using (\ref{eq_omega_q_simple}) for ungated graphene structure, the resonant freqency scales as $1/\sqrt{L}$ as opposed to scaling as $1/L$ for gate-voltage-controlled Fermi level. Thus, the resonant frequency for the same graphene patch length is lower for  gated graphene structure than for the ungated graphene structure. However, it must be noted that in either of the two cases, only a very small patch length of graphene would suffice for THz radiation. 

\vspace{0.35cm}
\section{Conclusions}
\vspace{0.15cm}
In this work, we propose device concepts in the terahertz frequency region using plasma oscillations in gated graphene structures. In such structures, it is shown that sufficiently long plasma waves exhibit a linear (sound-like) dispersion. Since graphene supports mass less Dirac fermions, the plasma waves in graphene can fall within the terahertz band even though their wavelength is sufficiently long. By modeling graphene as a resonant cavity or a voltage-controlled waveguide, different voltage-tunable terahertz devices can be materialized. We characterize the performance of on-chip interconnects and terahertz antennas built using mono- and multi-layer graphene patch on a substrate. Accounting for both intrinsic acoustic phonons and charged impurities scattering, we derive a simple scaling law between the applied gate voltage (``reconfigurable'' knob) and the thickness of the gate dielectric (``fixed'' upon manufacturing) that maximizes the propagation length of the plasma waves in both mono- and multi-layer graphene structures. Throughout the analysis, we consider capacitive effects in the device resulting from the quantum capacitance of graphene and the dangling bonds at the graphene-substrate interface. Further, apart from antennas and wave guides, graphene plasmons, via grated structures, can be profitably employed to dramatically increase the absorption of light at frequencies matched with the plasma resonant frequency. Hence, one can design extremely efficient terahertz photo-detectors with graphene heterostructures. 
\vspace{0.35cm}
\begin{appendices}
\appendix
\renewcommand \thesubsection{\Roman{subsection}}
\titlespacing\section{5pt}{12pt plus 4pt minus 2pt}{0pt plus 2pt minus 2pt}
\section{TM plasmon propagation mode in graphene from Maxwell's equation}
\begin{figure}[h!]
\centering
\includegraphics[width=3.0in]{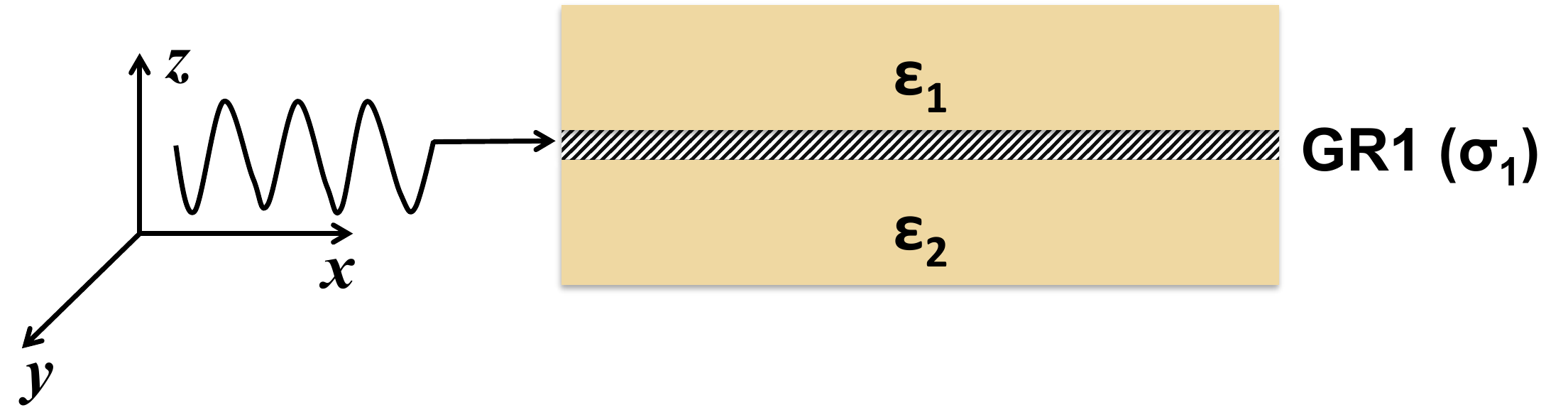} 
\caption{Schematic showing graphene embedded in a dielectric environment. The graphene layer is characterized by its dynamical conductivity, $\sigma_1$. The incoming light is assumed to be $p-$ polarized. That is, the electric field components are non-zero in $x-$ and $z-$ directions, while the magnetic field is non-zero in the $y-$ direction. Surface plasmons propagate along $x-$ direction.}
\label{fig_graphene_schematic_tm}
\end{figure}

\vspace{0.35cm}
Consider the graphene structure embedded between two distinct dielectric media characterized by their dielectric constants of $\epsilon_1$ and $\epsilon_2$ as noted in Fig. \ref{fig_graphene_schematic_tm}. 
Using  harmonic time dependence of the electromagnetic wave and separating out the $x-$, $y-$, and $z-$ components using Maxwell's equations, we arrive at the following set of equations. Note,  only $H_y$, $E_x$, and $E_z$ field components exist as we're considering TM propagation modes.

\begin{subequations}
\begin{equation}
\frac{\partial E_x}{\partial z}-i\beta E_z = i\omega\mu_0 H_y, 
\end{equation}
\begin{equation}
\frac{\partial H_y}{\partial z} = i\omega\epsilon_0\epsilon_r E_x, 
\end{equation}
\begin{equation}
i\beta H_y = -i\omega\epsilon_0\epsilon_r E_z.
\end{equation}
\end{subequations}

Further, $H_y$ must also satisfy the Helmholtz equation given as
\begin{eqnarray}
\frac{\partial^2 H_y}{\partial z^2}+(k_0^2\epsilon_r-\beta^2)H_y = 0,
\end{eqnarray} 
where $k_0$ is the free-space wave-vector equal to $\omega/c$, $\beta$ is the plasmon propagation constant.

For $z>0$, the following components of electric and magnetic fields are obtained
\begin{subequations}
\begin{equation}
H_y(z) = A_2 e^{i\beta x}e^{-k_2z}, 
\end{equation}
\begin{equation}
E_x(z) = \frac{iA_2}{\omega\epsilon_0\epsilon_2}k_2 e^{i\beta x}e^{-k_2 z},
\end{equation}
\begin{equation} 
E_z(z) = -\frac{-A_2\beta}{\omega\epsilon_0\epsilon_2}e^{i\beta x}e^{-k_2 z}.
\end{equation}
\end{subequations}

For $z<0$, the following components of electric and magnetic fields are obtained
\begin{subequations}
\begin{equation}
H_y(z) = A_1 e^{i\beta x}e^{k_1z}, 
\end{equation}
\begin{equation}
E_x(z) = -\frac{iA_1}{\omega\epsilon_0\epsilon_1}k_1 e^{i\beta x}e^{k_1 z}, 
\end{equation}
\begin{equation}
E_z(z) = -\frac{-A_1\beta}{\omega\epsilon_0\epsilon_1}e^{i\beta x}e^{k_1 z}.
\end{equation}
\end{subequations}

Now we must apply boundary conditions at $z=0$. The tangential component of $E$ must be matched. That is, $E_x(z=0^+) = E_x(z=0^-)$. Further, the discontinuity in the tangential component of $H$ at $z=0$ must be equal to $\vec{J}\times \hat{n}$, where $\vec{J}$ is the electric current and $\hat{n}$ is the unit vector perpendicular to the surface. Doing this, we obtain
\begin{subequations}
\begin{equation}
A_2 \frac{k_2}{\epsilon_2} = -A_1 \frac{k_1}{\epsilon_1}, \label{eq_tem1}
\end{equation}
\begin{equation}
A_2 - A_1 = -i\frac{\sigma(\omega,\beta)}{\omega\epsilon_0\epsilon_2}A_2k_2 \label{eq_tem2}.
\end{equation}
\end{subequations}

\noindent Using: 
$k_i^2 = \beta^2-k_0^2\epsilon_i$, where $i = 1,2$ denotes the region and (\ref{eq_tem1})-(\ref{eq_tem2}), we can show that the SPP dispersion relation for TM waves is given as
\begin{eqnarray}
\frac{\epsilon_1}{\sqrt{\beta^2-\frac{\epsilon_1\omega^2}{c^2}}}+\frac{\epsilon_2}{\sqrt{\beta^2-\frac{\epsilon_2\omega^2}{c^2}}} = -\frac{i\sigma(\omega,\beta)}{\omega\epsilon_0}.
\end{eqnarray} 

For the case when $\beta >> \frac{\omega}{c}$ (non-retarded regime), the TM plasmon dispersion relationship can be simplified to 
\begin{eqnarray}
\beta \approx \epsilon_0 \frac{\epsilon_1+\epsilon_2}{2}\frac{2i\omega}{\sigma(\omega,\beta)} \label{eq_spp_tm_nr}.
\end{eqnarray}

Again, separating out the real and imginary components of $\beta$ and $\sigma$ and equating the real and imaginary components of (\ref{eq_spp_tm_nr}), we can show
\begin{subequations}
\begin{equation}
\beta_R = \epsilon_0 \frac{\epsilon_1+\epsilon_2}{2}\left(\frac{2\omega\sigma_I}{\sigma_R^2+\sigma_I^2}\right), 
\end{equation}
\begin{equation}
\beta_I = \epsilon_0 \frac{\epsilon_1+\epsilon_2}{2}\left(\frac{2\omega\sigma_R}{\sigma_R^2+\sigma_I^2}\right).
\end{equation}
\end{subequations}

\begin{figure}
\centering 
\includegraphics[width=3in]{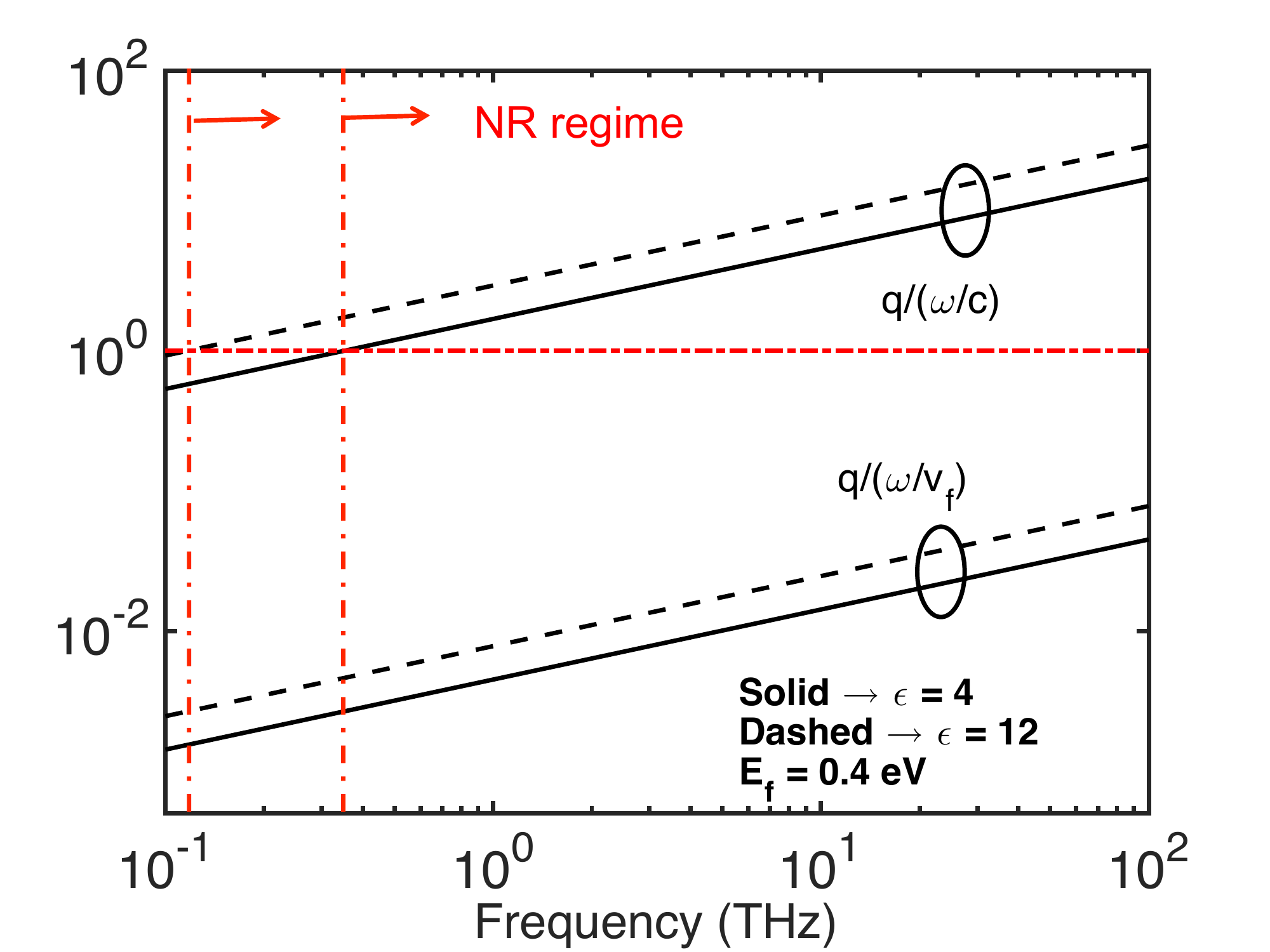} 
\caption{Range of the validity of (\ref{eq_omega_derived_tm}) for TM plasmon propagation.}
\label{fig_range_validity_TM_plasmon}
\end{figure}

Using the 2D conductivity of graphene from (\ref{eq_sigma_simple}) in Section II, $\beta_R$ can be related to the frequency $\omega$ and the Fermi level $E_f$ according to 
\begin{eqnarray}
\beta_R = \left(\frac{2\pi\hbar^2\epsilon_0\epsilon}{e^2|E_f|}\right)\omega^2,
\end{eqnarray} 
where $\epsilon = (\epsilon_1+\epsilon_2)/2$. Re-writing the above equation 
\begin{eqnarray}
\omega = \frac{1}{\hbar}\sqrt{\frac{e^2|E_f|\beta_R}{2\pi\epsilon_0\epsilon}}.
\label{eq_omega_derived_tm}
\end{eqnarray}
Note that the above equation is identical to (\ref{eq_omega_q_simple}) in Section II with $q$ standing for the real part of the propagation constant $\beta_R$. The range of validity of the above equation $\omega/c << q < \omega/v_f$ is demonstrated in Fig. \ref{fig_range_validity_TM_plasmon} for various values of $\epsilon$ at a Fermi level of $E_f$ = 0.4 eV. While the upper bound of $q < \omega/v_f$ is always satisfied in the plot, the lower bound of $q >> \omega/c$ for the validity of the NR regime depends on the material parameters $\epsilon$ and $E_f$.

\section{Electron relaxation rate in graphene}
\vspace{0.3cm}
\subsection{Scattering due to acoustic phonons}
\vspace{0.3cm}
The scattering time due to an elastic collision with acoustic phonon is defined as 
\begin{eqnarray}
\frac{1}{\tau_{\bf{k}}} = \Sigma_{\bf{k}}{\Gamma(\bf{k},\bf{k}')(1-\cos(\theta)},
\label{eq_tau_ph}
\end{eqnarray}
where $\theta$ is the scattering angle between $k$ and $k'$ states, and $\Gamma(\bf{k},\bf{k}')$ is the transition rate given as~\cite{stauber07}, \cite{hwang08}
\begin{eqnarray}
\Gamma(\bf{k},\bf{k}') = \frac{2\pi}{\hbar}|H_{\bf{k},\bf{k}'}|^2\delta(\hbar v_f k'-\hbar v_f k - \hbar \omega),
\label{eq_gamma_ph}
\end{eqnarray}
where $\hbar\omega$ is the phonon energy and $H_{\bf{k},\bf{k}'}$ is the matrix element for scattering by phonons. The matrix element for acoustic phonon scattering is given as 
\begin{eqnarray}
H_{\bf{k},\bf{k}'} = \cos(\theta/2)K_{\bf{q}}A_{\bf{q}}\delta_{\bf{k+q},\bf{k}'}e^{-i\omega t},
\label{eq_Hk_ph}
\end{eqnarray}
where $|K_{\bf{q}}|^2 = D_{A}^2q^2$ and $|A_{\bf{q}}|^2 = \hbar/(2\rho A\omega_{\bf{q}})N(\omega_{\bf{q}})$. Here $D_A$ is the electron acoustic deformation potential, estimated to be of the order of 3$t$, where $t$ is the nearest neighbor hopping integral, $N(\omega_{\bf{q}})$ is the occupation probability of phonons, and $\rho$ is the material density of graphene. Combining (\ref{eq_gamma_ph}) and (\ref{eq_Hk_ph}) with the conservation of momentum ($\bf{k+q} = \bf{k'}$) and inserting in (\ref{eq_tau_ph}), the final result for scattering time due to acoustic phonons is given as 
\begin{eqnarray}
\tau_k = \frac{4\hbar^2\rho v_s^2 v_f}{D_A^2k_BT}\frac{1}{k},
\label{eq_tauk_ac}
\end{eqnarray} 
where $v_s$ is the velocity of acoustic phonons. In the above equation it is assumed that the phonon occupation probability is $N_q \approx k_BT/(\hbar\omega_q)$. Further, in the case of graphene, scattering due to longitudinal acoustic (LA) phonon modes dominates as coupling of electron-phonon states for other phonon modes is too weak or the energy scales of the (optical) phonon modes are too high for the temperature range of interest. Equation (\ref{eq_tauk_ac}) can be re-written in terms of the energy $E_f$ according to 
\begin{eqnarray}
\tau_{ac} = \tau_{k_F} = \frac{4\hbar^3}{D_A^2}\frac{\rho (v_sv_{f})^2}{k_BT}\frac{1}{E_f},
\end{eqnarray} 
where $k_F$ is the Fermi wave-vector. The velocity of LA phonons in graphene  is 7.33$\times 10^3$ m/s. 
\vspace{0.35cm}
\subsection{Scattering due to charged impurities}
\vspace{0.35cm}
The scattering rate due to impurities is given as 
\begin{eqnarray}
\frac{1}{\tau_{\bf{k}}} = N_i \Sigma_{\bf{k'}}{\Gamma(\bf{k, k'})}(1-\cos(\theta_{\bf{k,k'}})),
\end{eqnarray}
where $N_i$ is the number of impurities and the transition from state $\bf{k}$ to $\bf{k'}$ is given by the Fermi Golden rule according to 
\begin{eqnarray}
\Gamma(\bf{k, k'}) = \frac{2\pi}{\hbar}|<\bf{k}|V_{scat}|\bf{k'}>|^2\delta(E_{\bf{k}}-E_{\bf{k'}}).
\end{eqnarray}
With the Fourier transform of the scattering potential $V_{scat}(q)$, the scattering rate can be written as 
\begin{eqnarray}
\frac{\hbar}{\tau_{k_F}} = \frac{n_i^{scat}}{8}\rho(E_f)\int{d\theta |V_{scat}(q)|^2}(1-\cos(\theta))^2,
\end{eqnarray}
where $n_{i}^{scat}$ is the impurities density of the scattering potential, $q = 2k_F\sin(\theta/2)$, and $\rho(E_f)$ is the 2D density of states in graphene at the Fermi level. Using the Thomas-Fermi scattering potential for long-range Coulomb potential (see Eq. (28) in \cite{stauber07}), the scattering time at $E_f >> k_BT$ (large doping) is given as \cite{stauber07}, \cite{hwang09} 
\begin{subequations}
\begin{equation}
\tau_{imp} = \tau_{k_f} = \frac{\hbar^2 v_f k_F}{u_0^2},
\end{equation}
\begin{equation}
u_0 = \frac{\sqrt{n_i^C}Ze^2}{4\epsilon_0\epsilon(1+\gamma)}, 
\end{equation}
\begin{equation}
\gamma = \frac{\rho(E_f)e^2}{2\epsilon_0\epsilon k_F},
\end{equation}
\end{subequations}
where $Ze$ is the net charge of the impurity atom, and $n_i^C$ denotes the concentration of charged impurities in the sample. Note that we have assumed the scattering to be elastic in nature.

\noindent The net scattering rate is computed using the Mattheissen's sum rule. That is, $\Gamma_{net} = \Gamma_{ac}+\Gamma_{imp} = \tau_{ac}\tau_{imp}/(\tau_{ac}+\tau_{imp})$. 

\section{Calculation of gate-voltage scaling law for maximizing plasmon propagation length }
\vspace{0.35cm}
Using definition of propagation length in (\ref{eq_propagation_length}) and simplifying the D.C. conductivity in the case when $E_f >> k_BT$ (see (\ref{eq_sigma_simple})), we obtain
\begin{eqnarray}
L_{prop} = \sqrt{\frac{e^2|E_f|d}{\pi\hbar^2\epsilon_0\epsilon}}\frac{1}{\Gamma}.
\label{eq_prop_len_simple}
\end{eqnarray} 
Here, $\Gamma$ = $\Gamma_{ac}+\Gamma_{imp}$, which are computed in Appendix B. The relaxation rates $\Gamma_{ac}$ and $\Gamma_{imp}$ exhibit inverse dependence on the Fermi level, $E_f$. That is,
\begin{subequations}
\begin{equation}
\Gamma_{imp} = \frac{a}{|E_f|}, 
\end{equation}
\begin{equation}
\Gamma_{ac} = b|E_f|, 
\end{equation}
\label{eq_gamma_simplified}
\end{subequations}
where $a$ and $b$ are constants given as 
\begin{eqnarray}
a &=& \frac{n_i^CZ^2e^4}{16\hbar\epsilon_0^2\epsilon^2}\left(1+\frac{e^2}{\pi\hbar v_f\epsilon_0\epsilon}\right)^{-2}, \\
b &=& \frac{D_A^2k_BT}{4\hbar^3\rho v_s^2v_f^2}.
\end{eqnarray}
Substituting the definitions from (\ref{eq_gamma_simplified}) in (\ref{eq_prop_len_simple}), we obtain
\begin{eqnarray}
L_{prop} = \sqrt{\frac{e^2d}{\pi\hbar^2\epsilon_0\epsilon}}\left[\frac{E_f^{3/2}}{a+bE_f^2}\right].
\end{eqnarray}
Noting that away from the Dirac point and neglecting interface trap capacitance, 
$E_f = \hbar  v_f \sqrt{\pi C_{ox}/e V_g'}$, where $C_{ox}$ is the gate oxide capacitance, and $V_{g}'$ is the difference in gate voltage and the Dirac point. Using this relationship between $E_f$ and $V_{g}'$, propagation length can be further simplified as
\begin{eqnarray}
L_{prop} = A \left[\frac{V_{g}'^{3/4}}{a+\frac{\beta}{d}V_g'}\right],
\end{eqnarray}
where $A$ is a constant independent of the gate voltage, and $\beta = b \left(\hbar v_f\right)^2 \pi \epsilon\epsilon_0/e$. To find the optimal value of $V_g'$ that maximizes the propagation length, 
\begin{eqnarray}
\frac{\partial L_{prop}}{\partial V_{g}'} = \frac{\partial}{\partial V_g'}\left[\frac{V_{g}'^{3/4}}{a+\frac{\beta}{d}V_g'}\right] = 0.
\end{eqnarray}
Carrying out the differentiation in the above equation, the optimal gate voltage is given as 
\begin{eqnarray}
V_{g,opt}' = {\frac{3a}{\beta}}d. 
\end{eqnarray}
Hence, one arrives at the linear scaling law between gate voltage and the effective oxide thickness to maximize the propagation length for plasmons in a gated graphene structure. It must be noted that parameters $a$ and $\beta$ are independent of the number of layers in the multi-layer stack. Hence, $V_{g,opt}'$ does not depend on $N_{layer}$ as noted in Section IV.

\end{appendices}


\end{document}